# Modeling Lead-Rubber Seismic Isolation Bearings Using the Unified Mechanics Theory

**H. Martin Hernandez Morales**[1,2]

[1]Department of Civil, Structural and Environmental Engineering, State University of New York at Buffalo, NY, USA

[2]School of Civil Engineering, National University of Engineering, Peru

## ABSTRACT

Lead-rubber seismic isolation bearings (LRB) have been installed in a number of essential and critical structures, like hospitals, universities and bridges, located in earthquake-prone areas. The purpose of using LRB is providing the structure with period lengthening and the capacity of dissipating a considerable amount of earthquake energy to mitigate the effects of strong ground motions. Therefore, studying the damage mechanics of this kind of devices is fundamental to understand and accurately describe their thermo-mechanical behavior, so that seismically isolated structures can be designed more safely.

Traditionally, up to this point, the hysteretic behavior of LBR has been modeled using 1) Newtonian mechanics and empirical curve fitting degradation functions, or 2) heat conduction theories and idealized bilinear curves which include degradation effects. The reason for using models that are essentially phenomenological or that contain some adjusted parameters is the fact that universal laws of motion by Newton lack the term to account for degradation and energy loss of a system. In this paper, the Unified Mechanics Theory – which integrates laws of Thermodynamics and Newtonian mechanics at ab-initio level – is used to model the force-displacement response of LRB. When Unified Mechanics Theory is used, there is no need for curve fitting techniques to describe the degradation behavior of the material. Damage or degradation at every point is calculated using entropy generation and the physics based fundamental equations of the material along the Thermodynamics State Index (TSI) axis.

A finite element model of a lead-rubber bearing was constructed in ABAQUS, where a user material subroutine UMAT was implemented to define the Unified Mechanics Theory equations and the viscoplastic constitutive model for lead. Finite element analysis results were compared with experimental test data.

# 1    INTRODUCTION

Since the pioneering work of W. H. Robinson [1], it is well-known that the hysteretic behavior of lead rubber bearings can be simply modelled as a bilinear curve, where the initial elastic stiffness – that depends mainly on lead - is approximately 10 times the post yield stiffness, which is governed by rubber. Consequently, LRB has been included in general and specialized dynamic analysis computer programs, such as SAP 2000 [2] and 3D-BASIS-ME-MB [3], respectively.

Phenomenological models have been proposed in order to represent LRB behavior more realistically; i.e. to account for damage. For example, Kikuchi and Aiken model [4] can capture the degradation of the stiffness associated with load history, recognizing it as a *complex phenomenon*. Constantinou et al [5,6] studied the effects of load history too, but the major contribution of their research is modeling strength degradation through lead core heating effects without performing lab tests to calibrate the model [5, 6, 7]. However, the dynamic equations that complement the heat-conduction based formulation include the bilinear hysteretic model suggested by Robinson [1].

Given its effectiveness to model damage suffered by lead rubber bearings due to temperature increase, other heat-conduction representations integrated with phenomenological or adjusted models have been created. For instance, Wake et al [8] combine the Kikuchi and Aiken model [4] and the finite volume method applied to heat-conduction analysis in order to describe strength degradation in LRB during long-duration earthquakes. Nevertheless, as stated at the end of [4]*, "Further work to develop analytical models of elastomeric bearings… based upon fundamental principles of computational continuum mechanics is required".* Therefore, it is fundamental to highlight the need for applying continuum damage mechanics to better describe the hysteretic behavior of LRB.

In this paper, the Unified Mechanics Theory [9-33], which was originally developed for microelectronic solder joints in multilayered electronics packages, was used to model the hysteretic thermo-mechanical behavior of a lead-core rubber seismic isolation bearing. Finite element analysis results were compared with lab test data taken from [5, 6]. This is the first attempt to model the mechanical behavior of LRB by means of applying the Unified Mechanics Theory [34].

A brief explanation of the complex behavior of lead and why it is used in elastomeric bearings is given in section 2. Likewise, the fundamentals of the Unified Mechanics Theory and its application to model lead are expounded in this part. In section 3, the finite element model of a LRB is described. Finally, the results obtained from ABAQUS simulations are compared with test data.

## 2    MODELING THE MECHANICAL BEHAVIOR OF LEAD

Lead-rubber isolation bearings are assumed to have the idealized hysteretic behavior shown in Figure 1. The principal parameters are the characteristic strength $Q_d$ – which is similar to the shear yield force $F_Y$ – and the post-elastic stiffness $K_d$. The value of the elastic stiffness $K_{el}$ is expected to be close to $10K_d$ [1], while the yield displacement $Y$ varies from 12 to 25 mm.

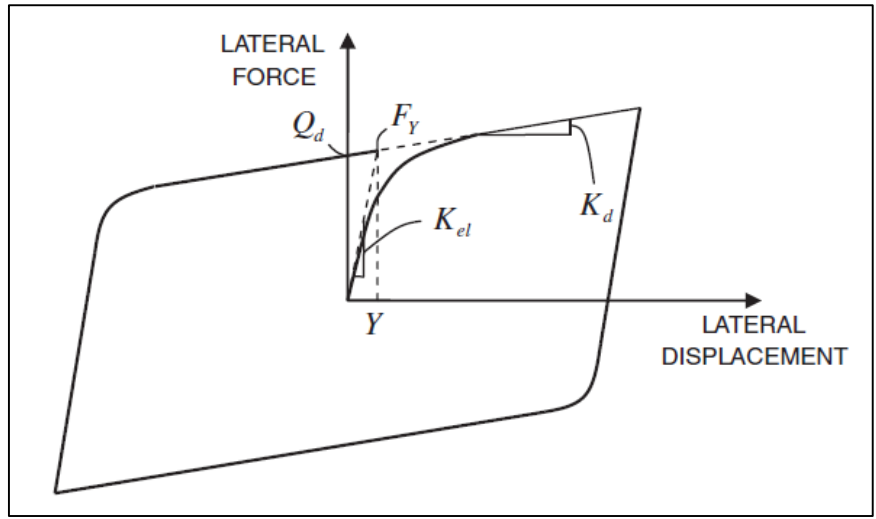

**Figure 1:** Idealized force-displacement response of a lead-rubber bearing [7].

The energy dissipation capacity of LRB is closely related to the characteristic strength, which is dependent on the cross section of the lead core, its confinement and its rate of shear strain deformation as well as the history of loading in terms of the number, amplitude and frequency of high-speed seismic motion cycles [6]. Thus, the mechanical behavior of lead strongly determines the main characteristics of the hysteretic behavior of lead-rubber bearings.

The mechanical properties of lead depend on multiple factors, such as crystal structure, grain size, degree of plastic deformation, temperature, purity, and strain rate. Lead has an A1-type Face-Centered Cubic (FCC) crystal structure, so that it remains ductile even at very low temperatures [6]. Although other metals like aluminum and copper have the same A1 lattice, plastically deforming lead at 20°C (or even at -4°C) is equivalent to plastically deforming those at 150°C and 200°C, respectively, because these temperatures represent the recrystallization temperature in each case. For steel, which constitutes plates and shims of LRB, and has an A2-type crystal structure, recrystallization temperature is even greater: 450°C [35].

Since the recrystallization temperature of lead is below room temperature, any deformation at or above it is "hot work", in which the processes of recovery, recrystallization and grain growth occur simultaneously [1, 35]. Deformation of polycrystalline metals results in elongation of the grains and a considerable increase in the number of defects (dislocations and vacancies) in each grain. Thus, during recovery, the stored energy of the deformed grains is reduced by the dislocations moving to form lower energy configurations, and by the annihilation of vacancies. Then, recrystallization occurs when deformed grains are replaced by small, new, undeformed grains that nucleate among the deformed ones and grow until them have been totally consumed. Further grain growth happens when some of the new grains become larger at the expense of others [6, 35]. Therefore, all of these mechanisms make lead more ductile even at low temperatures.

It is important to highlight that the greater the percentage of plastic deformation is, the lower the recrystallization temperature is. Thus, high plastic deformation and high temperatures facilitate recrystallization. Likewise, the beginning of recrystallization shifts to lower temperatures and less amount of plastic deformation on increasing the degree of purity of lead. In LRB, lead is of high purity; i.e., more than 99.9%. Moreover, the tensile strength of lead – and so the shear strength – drops with a decreasing strain rate because there is enough time for recrystallization to take place in case of low deformation rates, and it contributes to the inhibition of strain hardening.

As a conclusion, the mechanical properties of lead are continuously restored – not totally but largely – by the interrelated and simultaneous processes of recovery, recrystallization and grain growth, which makes lead an ideal material to be used in seismic isolators due to its good fatigue properties during cycling at plastic strains and high energy dissipation capacity, additionally to its relatively low shear yield stress of 10 to 12 MPa, and its elastic-viscoplastic behavior.

### 2.1 *Unified Mechanics Theory*

The Unified Mechanics Theory (UMT) unites laws of thermodynamics and Newton's universal laws of motion. This theory also modifies Newtonian space time coordinate system with the addition of the Thermodynamics State Index (TSI) axis, which goes between zero and one (see Figure 2). As a result of the new axis, derivative of displacement with respect to entropy is not zero as in classical Newtonian mechanics [9].

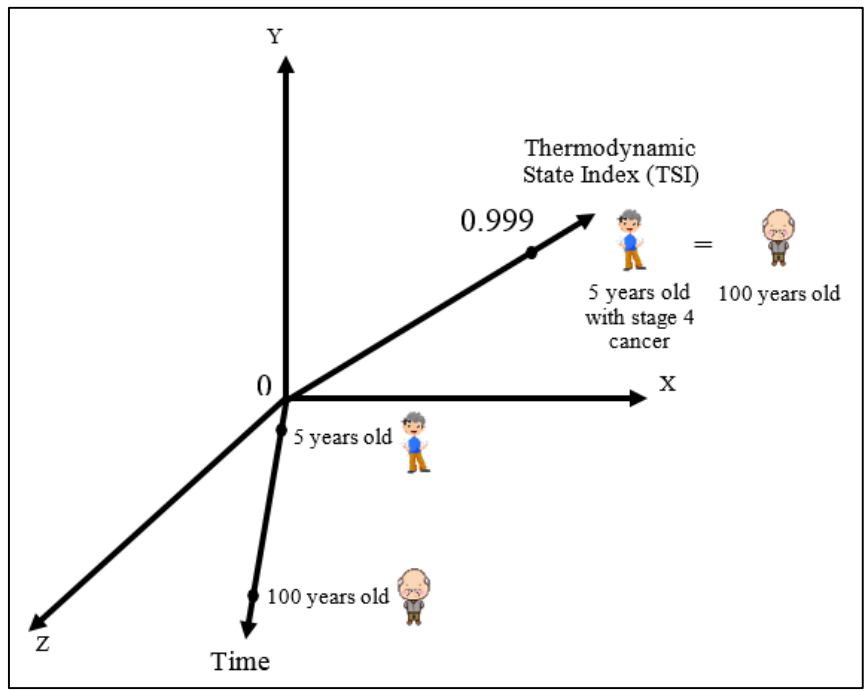

**Figure 2:** Dimensional Coordinate System in Unified Mechanics Theory [9, 10]

Degradation evolution occurring in a system, in this case LRB, follows the laws of thermodynamics. According to the second law of thermodynamics, entropy production rate becomes minimum when entropy is maximum. When a closed system can no longer generate entropy for a pre-defined mechanism, it is considered failed (or dead). Degradation happens due to irreversible entropy generation, which is always a non-negative quantity.

While Unified Mechanics Theory is not new, however, this is the first application to simulate the hysteretic behavior of LRB. Unified Mechanics Theory laws can be summarized in their simplified form as follows [9]:

$$\left(1 - \Phi(\dot{s}_f)\right) F dt = d(mv) \qquad (1a)$$

$$F = ku\left(1 - \Phi(\dot{s}_k)\right) \qquad (1b)$$

where: $F$ = force, $m$ = mass, $v$ = velocity, $k$ = stiffness, $u$ = displacement and $\Phi$ = Thermodynamic State Index, which is the normalized form of the second law of thermodynamics. $\dot{s}_f$ is the entropy generation responsible for the loss in momentum, while $\dot{s}_k$ is the entropy generation responsible for degradation of the stiffness. TSI is defined by the thermodynamic fundamental equation of the material which provides the entropy generation rate ($\dot{s}$) in the system for all active micro mechanisms. Value of TSI starts at zero and finally reaches one. In fact, equations (1a) and (1b) are nothing but the combination of second and third universal laws of motion of Newton and the second law of thermodynamics.

In this study, it is assumed that only lead provides entropy production in LRB. Of course, there is entropy generation in rubber layers and steel shims and plates. However, only lead experiences plastic deformation. Therefore, as a first iteration, it is considered that entropy production in rubber and steel is negligible compared to that in lead.

### Entropy balance

Entropy is a measure of how much energy is unavailable to do work as a result of loss and dissipation during a process in a closed system, and it is associated with a change in temperature [36]. Therefore, entropy can be created – in contrast to energy – and can only increase. The variation of total entropy $dS$ at a point may be written as the sum of two terms for a closed-isolated system as follows [37]:

$$dS = dS_e + dS_i \qquad (2)$$

where: $S_e$ = entropy gained from the transfer of heat from external sources across the boundary of the system, $S_i$ = entropy produced inside the system. The second law of Thermodynamics states that $dS_i \geq 0$, while $dS_e$ can be positive, zero or negative.

### *Degradation evolution function*

Degradation is the progressive deterioration that occurs in materials prior to failure. In Unified Mechanics Theory, degradation evolution function is based on the second law of thermodynamics and statistical physics, so that there is a relationship between irreversible entropy generation and degradation [9, 10, 37]. During the cumulative degradation process, the internal entropy production, which is a measure of disorder of a system, must increase. Thus, it can be used for mapping the evolution of degradation onto TSI axis. Thermodynamic State Index is given by [9, 10]:

$$\Phi = \Phi_{cr}\left[1 - e^{-\frac{m_s}{R}(s-s_0)}\right] \qquad (3)$$

where: $\Phi_{cr}$ = critical value of TSI (a user defined value), $m_s$ = molar mass, $R$ = universal gas constant, $s$ = entropy at a point at time t, and $s_o$ = initial entropy at a point at the beginning of the process (it can be taken as zero, for the reference state). The change in entropy per unit mass $(s - s_0)$ at a point can be calculated using the fundamental equation for a thermo-mechanical loading [10, 28]:

$$\Delta s = s - s_0 = \int_{t_0}^{t} \frac{\sigma : \dot{\varepsilon}^P}{T\rho} dt + \int_{t}^{t_0} \frac{k}{T^2 \rho} |grad\, T|^2 dt + \int_{t_0}^{t} \frac{r}{T} dt \qquad (4)$$

where: $\sigma$ = stress tensor, $\dot{\varepsilon}^p$ = plastic strain rate vector, $T$ = absolute temperature in Kelvin, $\rho$ = density, $k$ = thermal conductivity, and $r$ = distributed internal heat source per unit mass. Given that $\Delta s$ is a non-negative quantity, $\Phi \geq 0$ is always satisfied. Thermodynamic State Index (TSI = $\Phi$) is initially assumed to be zero, and it reaches the value of one when entropy generation rate is zero.

In this fundamental equation, it is assumed that entropy generation is limited to plastic work and internal heat generation under cyclic loading. In this study, for simplicity, only the first term of equation (4) was taken into account to calculate $\Delta s$. Therefore, certain amount of error is expected.

It must be emphasized that only irreversible entropy generation $S_i$ should be used as a basis for the systematic description of damage processes because $S_e$ has no influence in the degradation of materials [10, 37].

### 2.2 *Degradation coupled viscoplasticity material model*

Unified Mechanics Theory provides a framework for degradation of material properties according to laws of Thermodynamics. Thus, in accordance with the strain equivalence principle and Hooke's law, the elasticity constitutive relationship for lead, assuming small strains, can be written as:

$$d\sigma = (1 - \Phi)C_0(d\varepsilon - d\varepsilon^p - d\varepsilon^T) \qquad (5)$$

where: $C_0$ = initial stiffness matrix, $d\varepsilon$ = total strain increment, $d\varepsilon^p$ = inelastic strain increment, $d\varepsilon^T$ = thermal strain increment.

### Von Mises yield surface with isotropic and kinematic hardening

Knowing that lead has an elastic-viscoplastic behavior, Von Mises type yield function is used to separate elastic and inelastic response [25]:

$$F = \|S - X\| - \sqrt{\frac{2}{3}}\sigma_y = \sqrt{(S - X):(S - X)} - \sqrt{\frac{2}{3}}(\sigma_{y0} + R) \tag{6}$$

where: $F$ = yield surface separating the elastic from the inelastic response, $S$ = deviatoric component of the total stress tensor $\sigma$, $X$ = deviatoric part of the back stress tensor, $\sigma_{y0}$ = initial size of the yield surface, $R$ = isoparametric hardening function giving evolution of the size of the yield surface. $R$ and $\dot{X}$ can be calculated by [25, 38]:

$$R = R_\infty(1 - e^{-c\alpha}) \tag{7}$$

$$\dot{X} = \frac{2}{3}c_1\dot{\varepsilon}^P - c_2 X \dot{\alpha} \tag{8}$$

where: $R_\infty$ = isotropic hardening saturation value, $c$ = isotropic hardening material parameter, $c_1 = X_\infty$ = non-linear kinematic hardening saturation value, $c_2$ = nonlinear kinematic hardening material parameter, $\dot{\alpha}$ = equivalent plastic strain rate, which is given by:

$$\dot{\alpha} = \sqrt{\frac{2}{3}(\dot{\varepsilon}^p\dot{\varepsilon}^p)} \tag{9}$$

### Flow rule and consistency parameter

The evolution of the plastic strain vector is represented by a general flow rule of the form:

$$\dot{\varepsilon}^P = \dot{\gamma}\frac{\partial F}{\partial \sigma} = \dot{\gamma}\hat{n} \tag{10}$$

where: $\hat{n}$ = vector normal to the yield surface that specifies the direction of plastic flow, $\dot{\gamma}$ = rate of plastic flow. For a rate dependent material model, $\dot{\gamma}$ can be calculated by:

$$\dot{\gamma} = \frac{\langle \phi(F) \rangle}{\eta} \tag{11}$$

where: $\eta$ = viscosity material parameter, $\langle \ \ \rangle$ = Macaulay brackets, $\phi(F)$ = material specific function defining the character of the viscoplastic flow.

***Viscoplastic creep law***

The following creep law [21, 25, 30, 39] was used for lead:

$$\dot{\varepsilon}^P = \frac{AD_0Eb}{kT} \left( \frac{\langle F \rangle}{E} \right)^n \left( \frac{b}{d} \right)^p e^{-\frac{Q}{RT}} \frac{\partial F}{\partial \sigma} \tag{12}$$

where: $A$ = temperature and rate dependent dimensionless material parameter, $D_0$ = frequency factor, $E$ = temperature dependent Young's modulus, $b$ = characteristic length of crystal dislocation (magnitude of Burger's vector), $k$ = Boltzmann's constant, $n$ = stress exponent for viscoplastic deformation rate, $d$ = average grain size, $p$ = grain size exponent, $Q$ = creep activation energy.

Then, from equations (10), (11) and (12):

$$\langle \phi(F) \rangle = \langle F \rangle^n \tag{13}$$

$$\eta = \frac{kT}{AD_0E^{1-n}b} \left( \frac{d}{b} \right)^p e^{\frac{Q}{RT}} \tag{14}$$

The material model presented in this section was implemented in the general-purpose finite element program ABAQUS using UMAT subroutine option to model lead.

## 3    FINITE ELEMENT MODEL OF A LEAD-RUBBER BEARING

The lead-rubber bearing analyzed herein is shown in Figure 3. It is a large-scale LRB which was designed for a bridge application [5, 6]. The lead-core area is 0.025 m$^2$ (38.5 in$^2$) and the bonded rubber area is 0.650 m$^2$ (1007.9 in$^2$). Given that each rubber layer has a thickness of 0.01 m (0.4 in), the shape factor (S = loaded area / area free to bulge) is around 22, value associated with high vertical stiffness and adequate horizontal flexibility [40].

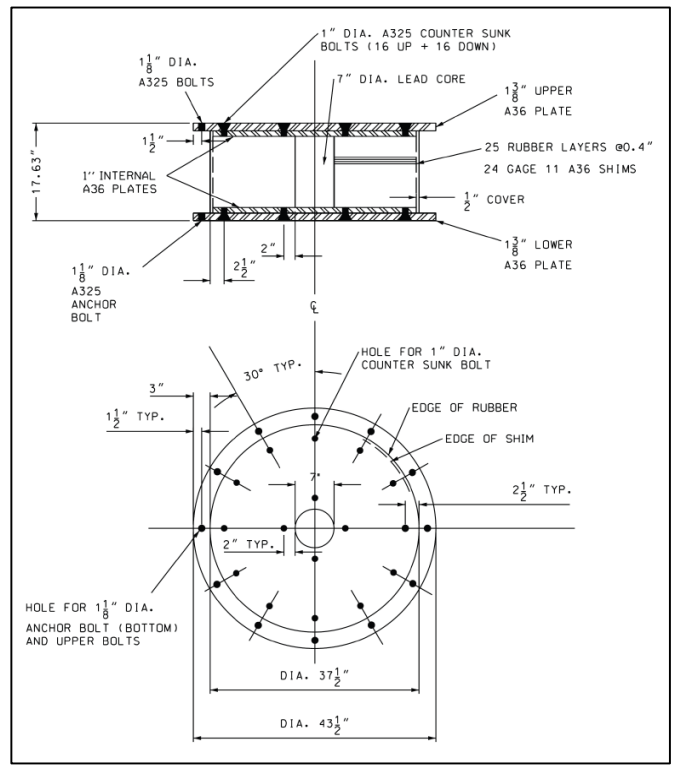

**Figure 3:** Geometry of the LRB seismic isolator under study [5, 6].

The bearing was tested by applying a constant vertical load of 3783 kN (850.5 kip) and a sinusoidal horizontal displacement with a period $T = 2s$ and amplitud $D_M = 0.305 m$ (12 in), which leads to a peak velocity $v_{max} = 0.958 \, m/s$ (37.7 in/s) and a maximum shear strain $\gamma_{max} = 120\%$ [5, 6]. This experiment was simulated in ABAQUS.

## 3.1 Properties of materials

Table 1 presents several properties of lead deemed in the finite element analysis of the LRB.

**Table 1:** Thermo-mechanical properties of lead.

| Property | Symbol | Value | Unit |
|---|---|---|---|
| Young's modulus | E | 1050-2.14T | MPa |
| Poisson's ratio | ν | 0.44 | |
| Shear yield stress | $\sigma_y$ | 23-0.039T | MPa |
| Kinematic hardening saturation value | $X_\infty$ | 510 | MPa |
| Isotropic hardening saturation value | $R_\infty$ | 9.37-0.019T | MPa |
| Density | ρ | 11.3 | t/m$^3$ |
| Molar mass | $m_s$ | 2.07E-04 | t/mol |
| Stress exponent | n | 1.67 | |
| Magnitude of Burger's vector | b | 3.18E-10 | m |
| Average grain size | d | 1.06E-05 | m |
| Grain size exponent | p | 3.34 | |
| Creep activation energy | Q | 44.7 | KJ/mol |

*T = Absolute Temperature (K)

Some mechanical properties of lead were obtained from experimental data provided by [5, 6]. For example, the Young's modulus $E = 0.42$ GPa at 20°C (Figure 4) was measured by means of a monotonic tensile test of a high-purity lead specimen at high strain rate (0.25/s). Although this value of $E$ does not belong to the commonly accepted range from 14.5 to 20 GPa [41], its validity can be easily verified through calculating the associated shear modulus and multiplying it by the shear strain obtained assuming an average value of the yield displacement $Y$ (Figure 1). The calculated shear yield stress $\sigma_Y$ is around 11.6 MPa, the initial value considered for the analysis [5]. Values of $E$ and $\sigma_Y$ at 20°C can be obtained by replacing T = 293.15K in the respective formulas of Table 1.

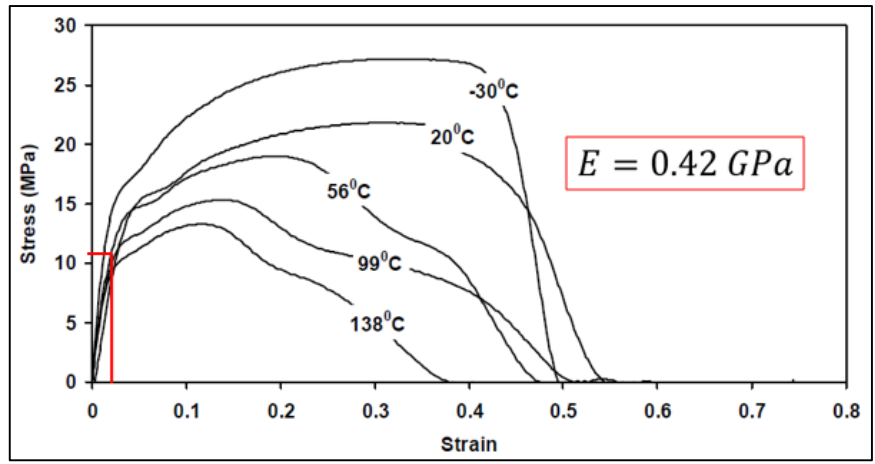

**Figure 4:** Determination of the Young's modulus $E$ of high-purity lead at 20°C [6].

However, the majority of lead material properties needed for the viscoplastic constitutive model were assumed to be similar to eutectic 63Sn/37Pb solder alloy [25]. This approximation was made because viscoplastic material properties of pure lead were not available. In order to verify the adequacy of such assumption, the monotonic test previously referred to was simulated in ABAQUS. Probable values of $R_\infty$ and $X_\infty$ parameters for high-purity lead were obtained (see Figure 5).

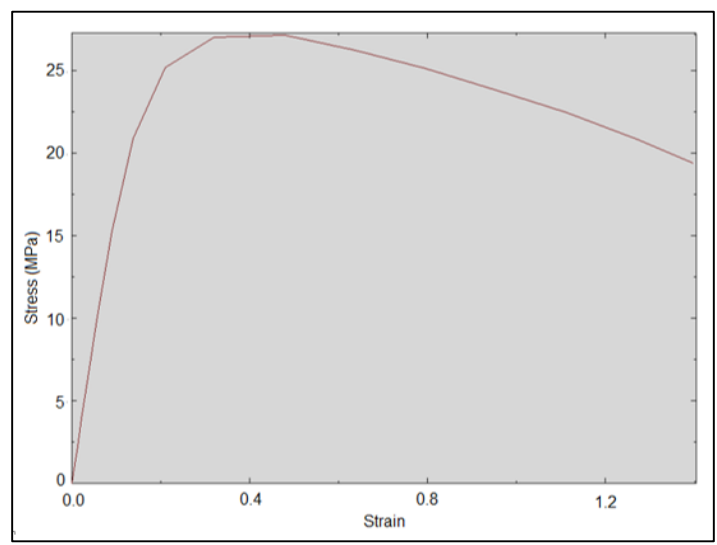

**Figure 5:** Simulation in ABAQUS of the monotonic tensile test (20°C) of Figure 4.

Rubber is a nearly incompressible material with Poisson's ratio $\nu$ very close to 0.5 (0.4998). Its shear modulus $G$ can be approximated to be 0.7 MPa for strains from 50 to 200%. Thus, the bulk modulus $K$ is near 2000 MPa, a value used in the design of elastomeric bearings [5]. In this case, a neo-Hookean material was selected to model rubber due to its simplicity – it is only necessary to know the values of bulk and shear modulus – and the lack of test data. Properties of $\nu = 0.495$, the ABAQUS's maximum allowable value, and $G = 0.7$ MPa were assigned. Moreover, steel was modeled as a linear-elastic material.

## 3.2    Analysis using ABAQUS and comparison with lab test data

Figure 6 shows the mathematical model used to simulate the hysteretic behavior of the LRB under study.

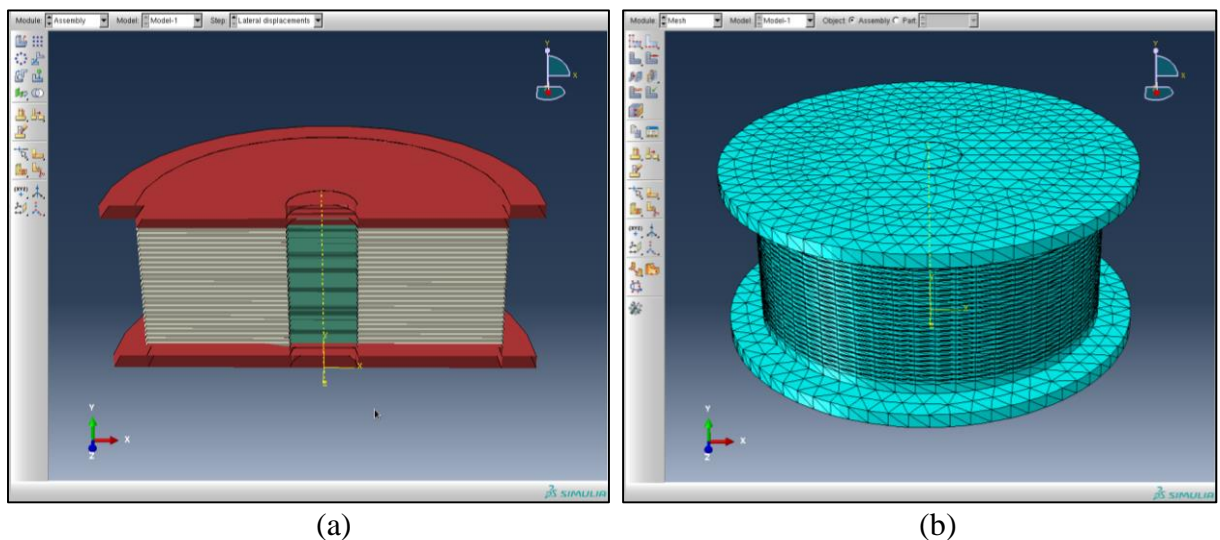

(a)                                        (b)

**Figure 6:** (a) LRB cross section and (b) mesh for finite element analysis in ABAQUS.

In Figure 6(a), internal and external upper and bottom plates, rubber layers, steel shims and the lead core of the isolator are shown. In Figure 6(b), the FE mesh of the lead rubber bearing model is depicted. The mesh has 217805 nodes and 103783 elements: 1007 linear wedge elements of type C3D6 for the lead core and 102776 quadratic tetrahedral elements of type C3D10 for the other parts of the LRB (see Figure 7).

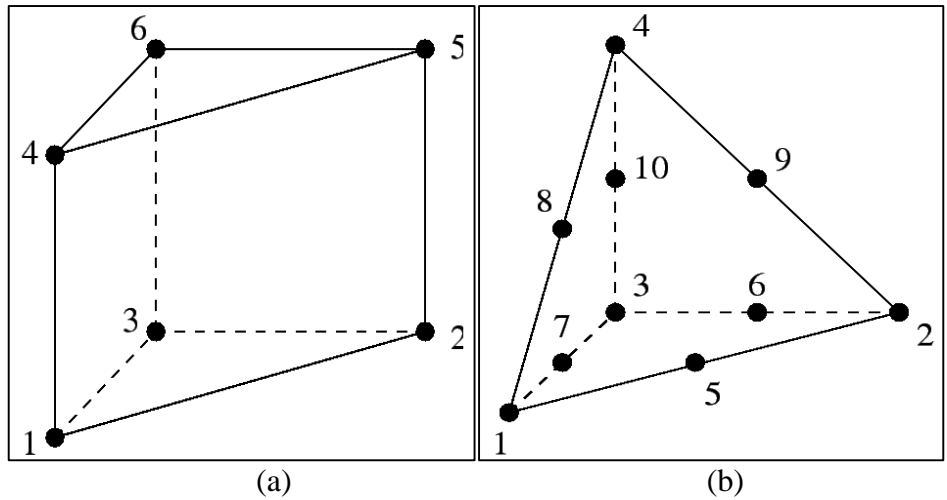

(a)  (b)

**Figure 7:** (a) 6-node wedge and (b) 10-node tetrahedral elements in ABAQUS [42].

It is important to remember that second and third term of the thermodynamic fundamental equation (4) were not considered in this analysis for simplicity. Results of the finite element heat transfer analysis of the isolator [6], which are shown in Figure 8, were imposed on the material properties.

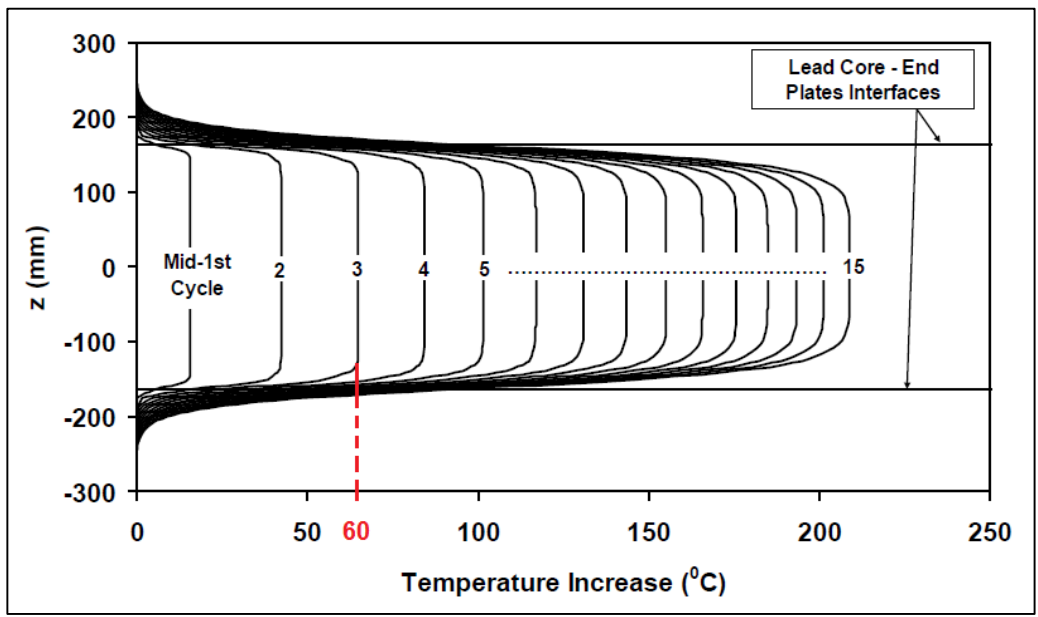

**Figure 8:** Vertical distribution of temperature increment at the center of the lead core as a function of the number of cycles [6].

Figure 9 shows the deformed LRB during the dynamic analysis in ABAQUS. Figure 10 presents the theoretical and experimental hysteresis loops of the lead-rubber bearing under study. Only the first three loops are computed by means of the Unified Mechanics Theory because the amount of energy dissipated by an isolator during a strong ground motion is approximately the same to that for 3 cycles at the maximum displacement $D_M$ [43].

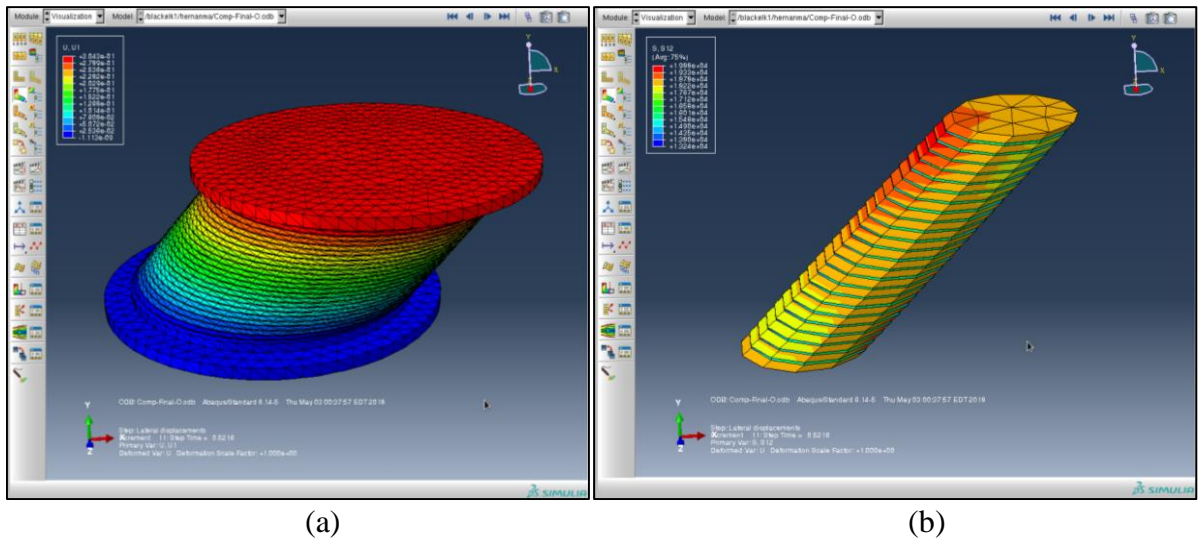

| (a) | (b) |

**Figure 9:** (a) LRB reaching the maximum displacement $D_M = 0.305m$, and (b) the respective shear stress distribution in the lead core.

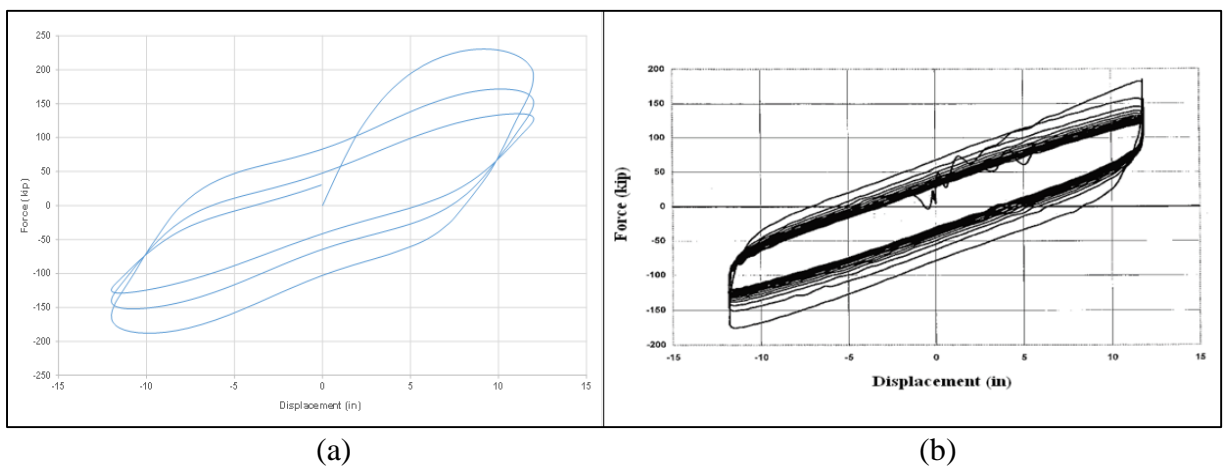

| (a) | (b) |

**Figure 10:** Force-displacement loops of the LRB from

(a) Unified Mechanics Theory and (b) lab test [5, 6].

Figure 10 demonstrates that Unified Mechanics Theory can properly simulate the hysteretic behavior of LRB – particularly, the second loop (see Figure 11) – although the match is not perfect. For example, the ascending branch of the first loop follows a different path than the one from lab test. Nevertheless, it should be noted the unexpected sinusoidal shape at the beginning of that loop which was imposed by the test machine. The part of the loops at reversal of motion could not be captured well too, as shown more clearly in Figure 11. However, that particular behavior probably could only be replicated by applying phenomelogical adjustments [7].

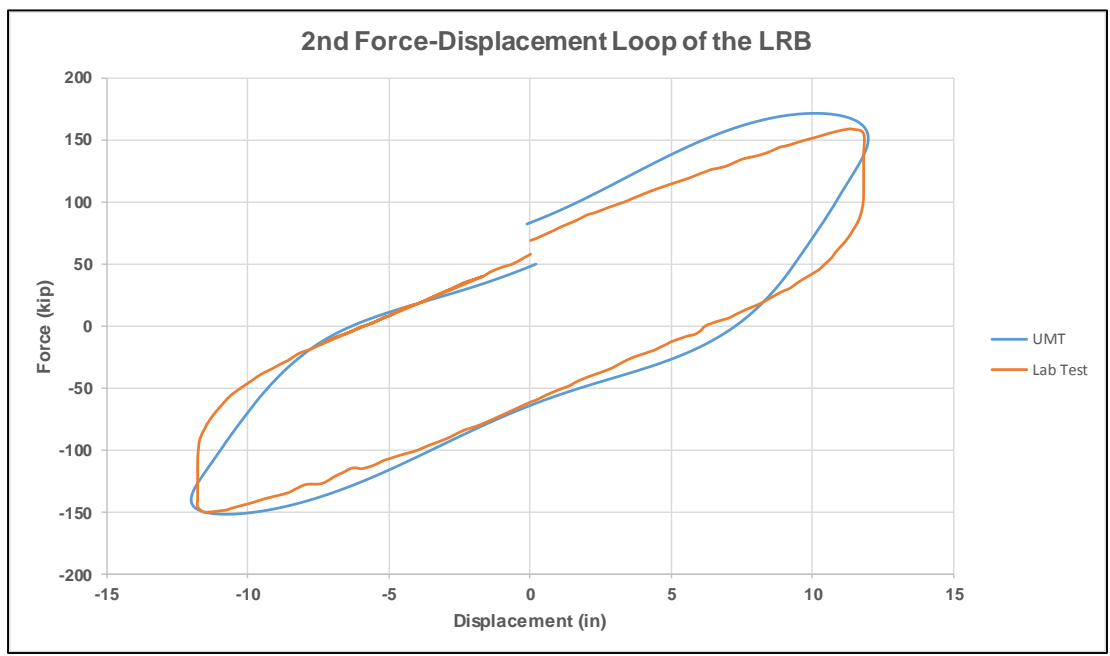

**Figure 11:** Comparison of the 2<sup>nd</sup> force-displacement hysteresis loop obtained from Unified Mechanics Theory and lab test [5, 6].

Tables 2 to 4 compare the energy dissipated per cycle ($EDC$), the effective stiffness ($K_{eff}$) and the effective damping ($\beta_{eff}$) obtained from Unified Mechanics Theory simulations and laboratory test.

**Table 2:** Comparison of the energy dissipated per cycle ($EDC$)

obtained from Unified Mechanics Theory and test data

| Cycle | EDC (kN-m) | | error (%) |
|---|---|---|---|
| | Test | Model | |
| **1** | 375.0 | 489.3 | 30.5 |
| **2** | **305.0** | **329.1** | **7.9** |
| **3** | 260.0 | 207.1 | -20.4 |

**Table 3:** Comparison of the effective stiffness ($K_{eff}$)

obtained from Unified Mechanics Theory and test data

| Cycle | Keff (kN/m) | | error (%) |
|---|---|---|---|
| | Test | Model | |
| **1** | 2583.0 | 2595.5 | 0.5 |
| **2** | **2240.1** | **2130.8** | **-4.9** |
| **3** | 2112.4 | 1836.3 | -13.1 |

**Table 4:** Comparison of the effective damping ($\beta_{eff}$)

obtained from Unified Mechanics Theory and test data

| Cycle | βeff | | error (%) |
|---|---|---|---|
| | Test | Model | |
| **1** | 0.249 | 0.323 | 29.9 |
| **2** | **0.233** | **0.265** | **13.4** |
| **3** | 0.211 | 0.193 | -8.4 |

## 4    CONCLUSIONS

This study is the first attempt in the literature to use the Unified Mechanics Theory to simulate the mechanical behavior of lead-rubber bearings. Degradation effects were included in the analysis through the Thermodynamic State Index (TSI); i.e., it was not necessary to apply empirical curve fitting degradation functions. As a result, a good match between analytical and experimental force-displacement curve was obtained, particularly for the second hysteresis loop.

Precisely, the percentages of error in energy dissipated per cycle, effective stiffness and effective damping are 7.9, -4.9 and 13.4%, respectively, for that loop. This is mainly important because second loops are often used to calculate nominal mechanical properties of seismic isolators. Besides, the model based on Unified Mechanics Theory could capture the significant difference in EDC between first and second – and first and third – hysteresis loops.

Even though the modified small-strain formulation implemented in ABAQUS by a UMAT subroutine demonstrated efficiency for modelling LRB isolators as a first order approximation, error between simulations and test data is partly due to this shortcoming. Therefore, large strain formulation must be included for a more precise finite element analysis. Other source of error is the use of material properties corresponding to eutectic 63Sn/37Pb solder alloys. Hence, it is necessary to obtain properties for high-purity lead (99.99%). Likewise, considering all mechanisms of equation (4) in the calculation of the entropy generation rate would yield more accurate results.